\documentclass[11pt]{article}   
\pdfoutput=1

\usepackage{amsthm,amsmath,amssymb,bbold,xcolor,sectsty,latexsym,mathtools,hyperref,mathrsfs,slashed,bm,epsfig,enumerate}

\usepackage[textwidth=6.5in,top=1.2in,bottom=1.2in]{geometry}

\newcommand{\beq}{\begin{eqnarray}} 
\newcommand{\eeq}{\end{eqnarray}}
\newcommand{\be}{\begin{equation}}
\newcommand{\ee}{\end{equation}}
\newcommand{\Mpl}{M_{\rm pl}}

\begin{document}

\thispagestyle{empty}
\begin{titlepage}
\nopagebreak

\title{ \begin{center}\bf Exploring the CMB Power Suppression\\ in Canonical Inflation Models \end{center} }

\vfill
\author{Mark Gonzalez$^{1}$\footnote{mark.gonzalez@tufts.edu}, \,\,\, Mark P.~Hertzberg$^{1,2,3,4}$\footnote{mark.hertzberg@tufts.edu}}

\date{\today}

\maketitle

\begin{center}
	\vspace{-0.7cm}
	{\it  $^{1}$Institute of Cosmology, Department of Physics and Astronomy,}\\
	{\it  Tufts University, Medford, MA 02155, USA}\\
	{\it $^2$School of Physics, University of New South Wales, Sydney, NSW 2052, Australia}\\
        {\it $^3$School of Physics, University of Sydney, Sydney, NSW 2006, Australia}\\
        {\it $^4$Department of Physics, Tokyo Institute of Technology,}\\
        {\it Ookayama, Meguro-ku, Tokyo 152-8551, Japan}

\end{center}

\bigskip

\begin{abstract}

There exists some evidence of a suppression in power in the CMB multipoles around $l\sim 20-30$. If taken seriously, this is in tension with the simplest inflationary models driven by a single scalar field with a standard type of slowly varying potential function $V(\phi)$. Such potential functions generate a nearly scale invariant spectrum and so they do not possess the requisite suppression in power. In this paper we explore if canonical two-derivative inflation models, with a step-like feature in the potential, can improve agreement with data. We find that improvement can be made when one utilizes the standard slow-roll approximation formula for the power spectrum. However, we find that in order to have a feature in the power spectrum that is sufficiently localized so as to not significantly disrupt higher $l$ or lower $l$, the potential's step-like feature must be so sharp that the standard slow-roll approximations break down. This leads us to perform an exact computation of the power spectrum by solving for the Bunch-Davies mode functions numerically. We find that the corresponding CMB multipoles do not provide a good agreement with the data. We conclude that, unless there is fine-tuning, canonical inflation models do not fit this suppression in the data.

\end{abstract}

\end{titlepage}

\setcounter{page}{2}

\tableofcontents


\section{Introduction}

The cosmic microwave background (CMB) provides the clearest information we have about the details of the early universe \cite{Hinshaw:2012aka,Ade:2015xua}. Its approximate isotropy can be accommodated beautifully within the theory of cosmic inflation \cite{Guth:1980zm,Linde:1981mu}. Furthermore, inflation accounts for the approximate homogeneity and fluctuations in the large scale structure of the universe. Detailed measurements are broadly in agreement with the predictions of the simplest inflationary models; namely that of a nearly scale invariant spectrum of fluctuations, with a small red tilt, adiabatic and Gaussian fluctuations, and a universe that is spatially flat.

However, recent observations of the CMB suggest that the fluctuations on a range of scales deviate from scale invariance in an interesting way. In particular if one decomposes the CMB into multipole moments $\mathcal{D}_l^{TT}$ and looks at the measured power in each multipole, one finds that while there is broad agreement with the predictions of inflation, there is a breaking of scale invariance seen in the appearance of a suppression in power in multipoles around $l\sim 20-30$ \cite{Ade:2015lrj} (see the data in Figs.~\ref{MultipolePlotApprox} \& \ref{MultipolePlotExact}). Other works have focussed on a possible overall suppression in power for {\em all} low-$l$ modes, which may be possible in open inflation \cite{White:2014aua}, but this will not be our focus here.

Now it is entirely possible that this suppression for $l\sim 20-30$ is just a statistical fluke, since inflation is a statistical theory built on the principles of quantum mechanics. So some variability in the measured power on various scales is expected, and since these scales are rather large, there is appreciable cosmic variance. Moreover, one should always be careful with regards to the ``look elsewhere effect", whereby one can always data mine and find anomalies if one looks at the data in certain ways; the significance is often reduced when a global statistical analysis is performed. However, it is at least plausible that this effect should be taken seriously, and suggests that the simplest inflationary models, which predict a nearly scale invariant spectrum, are in tension with the latest data. 

In general, however, it is difficult to definitively rule out the idea of inflation, since the theory is at such high energies and at very large field values. In this regime, the rules of relativity and quantum mechanics allow for a large number of operators to be relevant in controlling the physics. This is to be contrasted to the situation at low energies when we expand around the vacuum. In particular, if we consider a scalar field's self interactions of the form $V(\phi)=\sum_n c_n\, \phi^{n}/\Lambda^{n-4}$, one can usually be confident that the higher order terms in this expansion are irrelevant as they are suppressed by some large mass scale $\Lambda$ (we are working in 3+1 dimensions). However, in the inflationary phase, the inflaton $\phi$ is usually at such large field values that the entire tower of terms may be important. It has sometimes been suggested that this is the trouble with inflation; that it is not especially predictive. But this misses the target. The trouble is that the rules of quantum mechanics and relativity themselves permit this tower of terms and so they have somewhat limited predictive power in this particular extreme regime (while obviously being amazingly predictive in other regimes); inflation is simply a particular phase that is plausibly allowed by these over-arching principles.

Moreover, there appears to be tremendous freedom allowed by relativity and quantum mechanics on the number of scalar fields and on many other types of operators, namely higher derivative interactions, such as $\sim(\partial\phi)^4$. However this is where the general principles of effective field theory provide a great amount of guidance. If one is in a regime in which such higher derivative terms are important, then one is somewhat near the cutoff of the effective theory. While it is possible to tune parameters to be in a regime in which (a) these higher derivative terms are important and (b) the effective field theory is still valid, it generally requires special pleading of parameters. Furthermore, such models are not suggested by any existing data, as such scenarios tend to predict large non-Gaussianity, while the data is consistent with Gaussianity. Also, while it is interesting to study multi-field models (an interesting example is Ref.~\cite{Blanco-Pillado:2015bha}), in the absence of special pleading, inflation usually leads to single field attractor behavior. 

All together, the class of models with a single scalar field governed by a standard two-derivative action, are by far the most well motivated from the general considerations of effective field theory. In these models the only residual freedom is the potential function $V(\phi)$ mentioned above, leaving a single functional freedom in the theory; these are the ``canonical inflation models".

In this paper, we focus on these canonical inflation models, and explore if some potential function $V(\phi)$ can accommodate the suppression in power on scales $l\sim 20-30$. To do so, one must introduce a feature into the potential so as to create a dip in power on just the right set of scales. It is not clear what the microphysics underlying such a feature would be, but it is allowed in principle, so long as the feature isn't so sharp that particle scattering is in conflict with unitarity. 

A suppression in power can be obtained from a steep potential. Furthermore, in order to not disrupt the nearly scale invariant spectrum for $l\ll 20$ and $l\gg 30$, we need this steep part of the potential to be localized. Hence this leads us to consider a class of potential functions that have a step-like feature, with a height, width, and location that we shall take as adjustable parameters (related work includes Refs.~\cite{Novaes:2015uza,Mooij:2015cxa,Qureshi:2016pjy} and references therein). We find that with these potentials and using the slow-roll approximations to obtain the power spectrum we can obtain moderate improvement in the data compared to standard potentials without this feature. However, we find that in order to improve the fit to data, the width of the step needs to be so small that the standard slow-roll approximations break down. We then perform an exact analysis by solving for the Bunch-Davies mode functions numerically. We then find that the power spectra exhibit significant oscillations, which does not fit the data well. We conclude that in order to obtain agreement with data, one needs a highly fine-tuned potential that has a range of features that conspire to remove these prominent oscillations.

Our paper is organized as follows: In Section \ref{Models} we outline the class of models we study. In Section \ref{PowerSpectrum} we layout the computation of power spectra both in an approximate slow-roll form and an exact form. In Section \ref{CMB} we illustrate the predictions for the CMB multipole moments. In Section \ref{Statistics} we perform a systematic exploration of the fit to data, by computing the sum of squared differences between theory and data. Finally, in Section \ref{Discussion} we discuss our results.

\section{Class of Models}\label{Models}

As explained above, the most well motivated model for inflation, based on the general principles of effective field theory, is the standard two-derivative action for gravity with a single scalar field. In this case we can always exploit field re-definitions of the metric $g_{\mu\nu}$ to go to Einstein frame and field redefinitions of the scalar $\phi$ to make the kinetic term canonical. This gives the following action (signature $+---$ and units $\hbar=c=1$)
\be
 S = \int d^4 x \,\sqrt{-g} \left[ {1\over2}\Mpl^2\mathcal{R} + \frac{1}{2} \partial_\mu \phi \partial^\mu \phi - V(\phi) \right],
\ee
where $\mathcal{R}$ is the Ricci scalar and $\Mpl\equiv1/\sqrt{8\pi G_N}$ is the reduced Planck mass.

In order to specify a class of models, we would like to begin with a standard type of potential function $V_0(\phi)$ that gives rise to a nearly scale invariant spectrum, as this is in rough agreement with data. We then deform the potential by introducing a step-like feature into it $\delta V(\phi)$ to try to obtain the requisite suppression in power on the appropriate scales. The total potential is then
\be
V(\phi)=V_0(\phi)+\delta V\!(\phi).
\ee 

A useful starting point is the simplest inflationary potential, namely a quadratic potential \cite{Linde:1983gd}
\be
V_0(\phi)={1\over2}m^2\phi^2. 
\label{V0}\ee
The overall measured amplitude of the variance in fluctuations, $A\approx 2.2 \times 10^{-9}$, can be accommodated by choosing the mass to be $m\approx 1.6\times 10^{13}$\,GeV. This model predicts a spectral index of $n_s\approx1-2/N_e$, where $N_e$ is the number of e-foldings of inflation. If we take $N_e\sim 50-60$, this predicts a spectral index of $n_s\sim0.96$, which is in good agreement with data. On the other hand, this model predicts a tensor-to-scalar ratio of $r\approx8/N_e$, giving $r\sim 0.15$, which is slightly higher than existing bounds on primordial B-modes. Hence one should more realistically move to models with an overall flatter potential to avoid over-production of gravitational waves. Nevertheless this model is sufficiently simple that it is useful to illustrate the basic idea, and we will make use of it. It is simple to generalize our method to flatter potentials.

\begin{figure}[t]
\centering
\includegraphics[width=10cm]{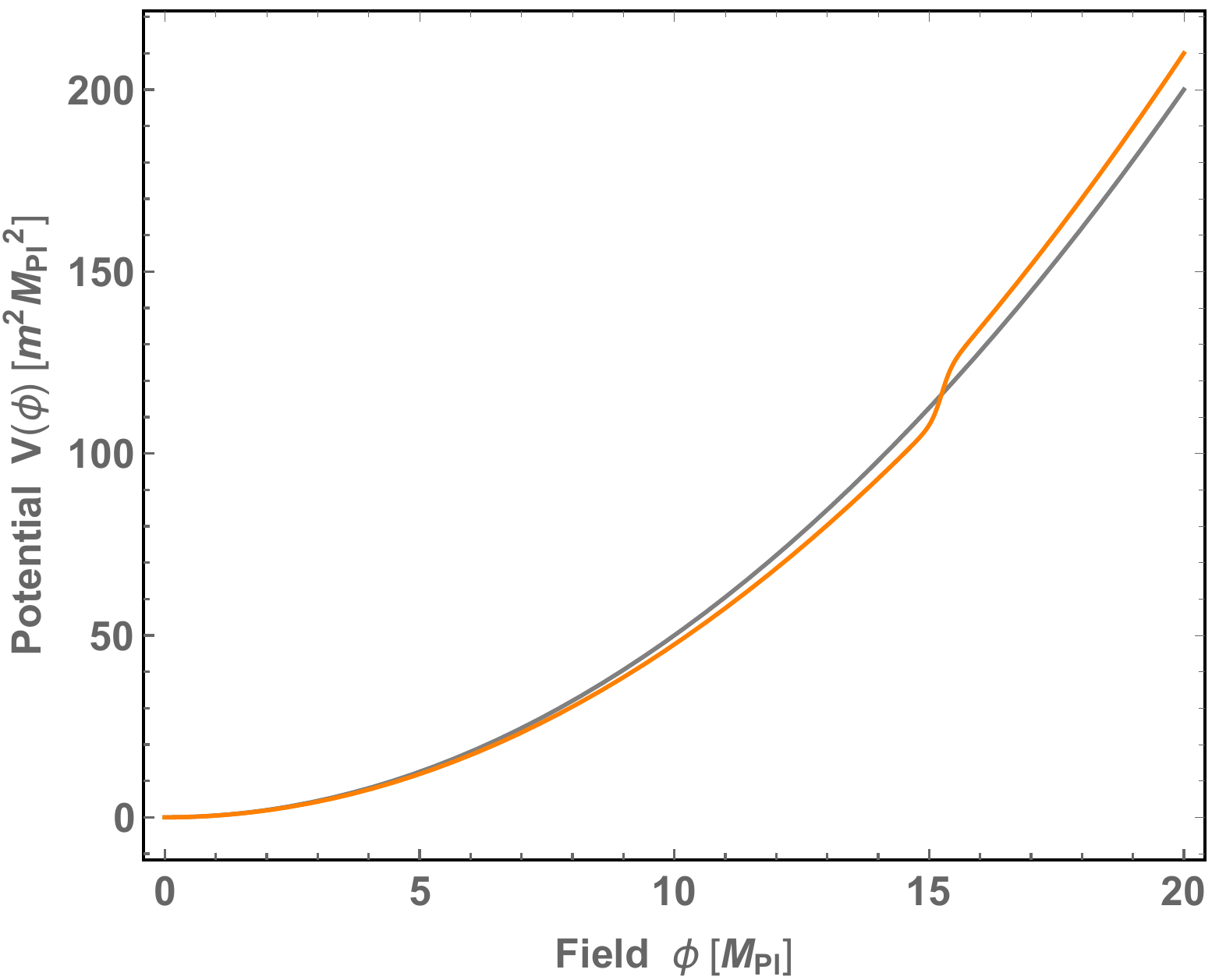}
\caption{A representation of the kind of potentials $V(\phi)$ considered in this paper. The gray curve is a standard quadratic potential $V_0(\phi)$, while the orange curve is the full potential including a step-like feature $\delta V\!(\phi)$, with $\alpha=0.05$ (though we will often focus on smaller $\alpha$ values in the remainder of the paper), $\gamma=5$, and $\phi^*=15.243\,\Mpl$.}
\label{Potential}
\end{figure}

In order to introduce a feature into the potential we would like to add a piece $\delta V$ that is localized around some special value $\phi^*$, leaving the potential of the quadratic form for $\phi\ll\phi^*$ and $\phi\gg\phi^*$. In between we would like to add some step-like function that has an amplitude and width that are adjustable parameters. The idea is that the feature introduces a localized steepness into the potential. This is important because (at least within the slow-roll approximation) the power is inversely proportional to $\epsilon_1$ (see ahead to Eq.~(\ref{Papprox})) and a steep potential has a larger $\epsilon_1$ and therefore a suppression in power. The specific form of the potential is not too important, just so long as it has these qualitative features. But for concreteness we choose the following functional form which has all of these properties
\be
\delta V\!(\phi) = \alpha\,\mbox{tanh}\!\left[\gamma\,(\phi-\phi^*)\over\Mpl\right] V_0(\phi),
\label{deltaV}\ee
where $\alpha$ is a dimensionless measure of its amplitude and $\gamma$ is a dimensionless measure of its (inverse) width. A plot of this form of the potential is given in Fig.~\ref{Potential}. We have checked that our basic conclusions extends to any qualitatively similar set of potential functions. For $\gamma\lesssim 1$, the UV cutoff on the effective theory is the Planck scale as usual. However, we will be interested in relatively sharp step-like features and so we will be exploring $\gamma \gg 1$. In this case the UV cutoff on the effective theory can be below the Planck scale. If one series expands the above potential, one finds that higher order operators are suppressed by a factor of $\Mpl/\gamma$ (times an overall factor involving $m^2$ and $\alpha$), which suggest that higher energy scattering of particles may violate the unitarity bound at the scale
\be
\Lambda_{UV}\sim{\Mpl\over\gamma}.
\ee
which acts as the cutoff on the effective theory. However we will not study $\gamma$ that are extremely large. The maximum we will explore is $\gamma\sim 100$. So the cutoff will remain  higher than both the Hubble scale during inflation and the energy density of inflation to the one-quarter power. Thus the effective theory can be used.

For $\phi\ll\phi^*$, tanh$[\gamma(\phi-\phi^*)/\Mpl]\to-1$, and the total potential becomes $V(\phi)\approx m^2(1-\alpha)\phi^2/2$. For $\phi\gg\phi^*$, tanh$[\gamma(\phi-\phi^*)/\Mpl]\to1$, and the total potential becomes $V(\phi)\approx m^2(1+\alpha)\phi^2/2$. By taking $\alpha\ll 1$ we ensure that the overall effective mass is not shifted significantly between the $\phi\ll\phi^*$ and the $\phi\gg\phi^*$ regimes. Hence it will still be the case that $m\approx 1.6\times 10^{13}$\,GeV to approximately match the overall measured amplitude of the variance in fluctuations. However, small adjustments in the value of $m$ will be made in order to provide an optimal fit to the data. 

\section{Power Spectra}\label{PowerSpectrum}

Our goal is to compute the spectrum of fluctuations in this class of models and compare to the latest CMB observations from Planck. In order to do so we need to first compute the homogeneous background evolution of the field $\phi_b$ and metric, and then compute the first order fluctuations.

The homogeneous background evolution is straightforward. In a spatially flat universe, the corresponding Friedmann and field equations are as usual
\beq
 &&\ddot{\phi}_b + 3 H \dot{\phi}_b + V'(\phi_b) = 0 ,
 \label{EOM}
 \\
 &&H^2 = \frac{1}{3\,\Mpl^2} \left( \frac{1}{2} \dot{\phi}_b^2 + V(\phi_b) \right), 
 \label{Friedmann}
\eeq
where $H=\dot{a}/a$ is the Hubble parameter, with $a$ the scale factor, and dots indicate derivatives with respect to time.

\begin{figure}[t]
\centering
\includegraphics[width=8cm,angle=-90]{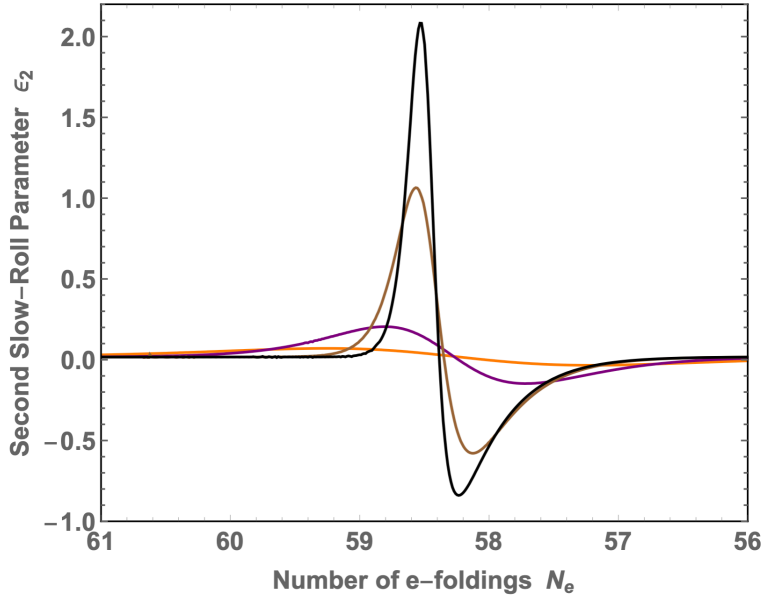}
\caption{The second slow-roll parameter $\epsilon_2$ versus the number of e-foldings $N_e$ until inflation ends, with $\alpha=0.0015$, $\phi^*=15.243\,\Mpl$, and for different values of the inverse width parameter; orange is $\gamma=5$, purple is $\gamma=10$, brown is $\gamma=30$, and black is $\gamma=50$.}
\label{SlowRoll}
\end{figure}

A phase of acceleration occurs when the following first slow-roll parameter $\epsilon_1$ is less than 1
\be
\epsilon_1 \equiv - \frac{\dot{H}}{H^2}. 
\label{epsilon}
\ee
In all our parameter searches, we ensure that $\epsilon_1\ll 1$ is satisfied. The corresponding number of e-foldings until inflation ends is determined by the following integral
\be
N_e (t) = \int_t^{t_{\rm end}} H (t)\, d t ,
\label{Ne}
\ee
where $t$ is the time at which one is computing the number of e-foldings and $t_{\rm end}$ is the time at which inflation ends. Furthermore, a condition to trust the standard slow-roll approximations is that the following second slow-roll parameter is also small
\be
\epsilon_2\equiv {\dot\epsilon_1\over H\,\epsilon_1}.
\ee
In Fig.~\ref{SlowRoll} we plot $\epsilon_2$ over time and see that for sufficiently large $\gamma$, $\epsilon_2$ becomes large when the field passes through the step-like feature in the potential.

\subsection{Density Perturbations}

We are interested in computing scalar perturbations around the homogeneous background (a related analysis can be carried out for tensor perturbations). Scalar perturbations arise in the scalar field $\delta\phi$ and the metric $\delta g_{\mu\nu}$ in a gauge dependent fashion. In the linear theory there is a useful gauge independent quantity which measures the curvature perturbation $\zeta$ and has the convenient feature that it is frozen on super-horizon scales. 

In the Gaussian approximation, the fluctuations in $\zeta$ are entirely characterized by its two-point correlation function. Statistical translation and rotation invariance implies that it can be described by a single function of one variable, the power spectrum ${\cal P}_\zeta (k)$. It is defined implicitly by
\be
\langle \hat{\zeta} ({\bf k}) \, \hat{\zeta} ({\bf k}') \rangle = (2\pi)^3 \delta^3 ({\bf k} + {\bf k}') {2 \pi^2 {\cal P}_\zeta (k)\over k^3},
\ee
where we have chosen to scale out a factor of $1/k^3$ for convenience. Note that this leaves ${\cal P}_\zeta(k)$ dimensionless, and if it were independent of $k$, it would correspond to a scale-invariant spectrum. Here $k=|{\bf k}|$ is the magnitude of the wavevector and is defined in comoving co-ordinates.

At the linear level there is an exact way to determine the power spectrum. We decompose the quantum field in terms of mode functions $v_k$ (the Mukhanov-Sasaki variable \cite{Sasaki:1986hm,Mukhanov:1988jd}). This is related to the curvature perturbation mode functions by $\zeta_k=v_k/(a\sqrt{2\,\epsilon_1})$, which obeys the following second order equation of motion
\be
 \frac{d^2 \zeta_k}{d \eta^2} + (2+\epsilon_2)a H{d\zeta_k\over d\eta}+k^2\zeta_k = 0,
\label{MSeq}\ee
where $\eta\equiv\int dt/a(t)$ is conformal time. To fully specify the mode function, we must impose boundary conditions. In order for each mode to begin sub-horizon in the usual Minkowski space vacuum, we impose the following boundary condition
\be
 v_k \to \frac{e^{-i k \eta}}{\sqrt{2k}},
\ee
in the distant past; this defines the so-called Bunch-Davies vacuum. The corresponding ground state wave-functional is a Gaussian with a variance that specifies the power spectrum ${\cal P}_\zeta$. It is straightforward to show that this is related to the square of the mode functions as
\be
 \mathcal{P_\zeta} (k) = \frac{k^3 |\zeta_k|^2}{2 \pi^2\Mpl^2}.
\label{Pexact}\ee

Now an important approximation emerges if the background always exhibits standard slow-roll inflation in quasi de Sitter phase, where the Hubble parameter and its derivatives change very slowly in time, implying $\epsilon_2\ll 1$. By solving Eq.~(\ref{MSeq}) in this slow-roll limit, one finds that the mode functions approach $|\zeta_k|\to H_k/(2\sqrt{k^3\,\epsilon_{1,k}})$, where $H_k$ and $\epsilon_{1,k}$ (the Hubble parameter factor and first slow-roll parameter, respectively) are evaluated at the time when the corresponding mode $k$ crosses the horizon, i.e., when $N_e=\ln(k_{\rm end}/k)$, where $k_{\rm end}$ is the scale that leaves the horizon at $t=t_{\rm end}$. This gives the following well known slow-roll approximation to the power spectrum
\be
 \mathcal{P_\zeta} (k) \approx \frac{H_k^2}{8 \pi^2\Mpl^2 \epsilon_{1,k}}.
\label{Papprox}\ee

In this work we will compare the results of using this approximate form eq.~(\ref{Papprox}), which is often sufficient in describing the predictions of inflation, to the exact form eq.~(\ref{Pexact}). In the case of our potential with a step-like feature it is possible to be in the regime in which this approximate form will be inaccurate as we discuss below.

\subsection{Numerical Results}

\begin{figure}[t!]
\centering
\includegraphics[width=8cm]{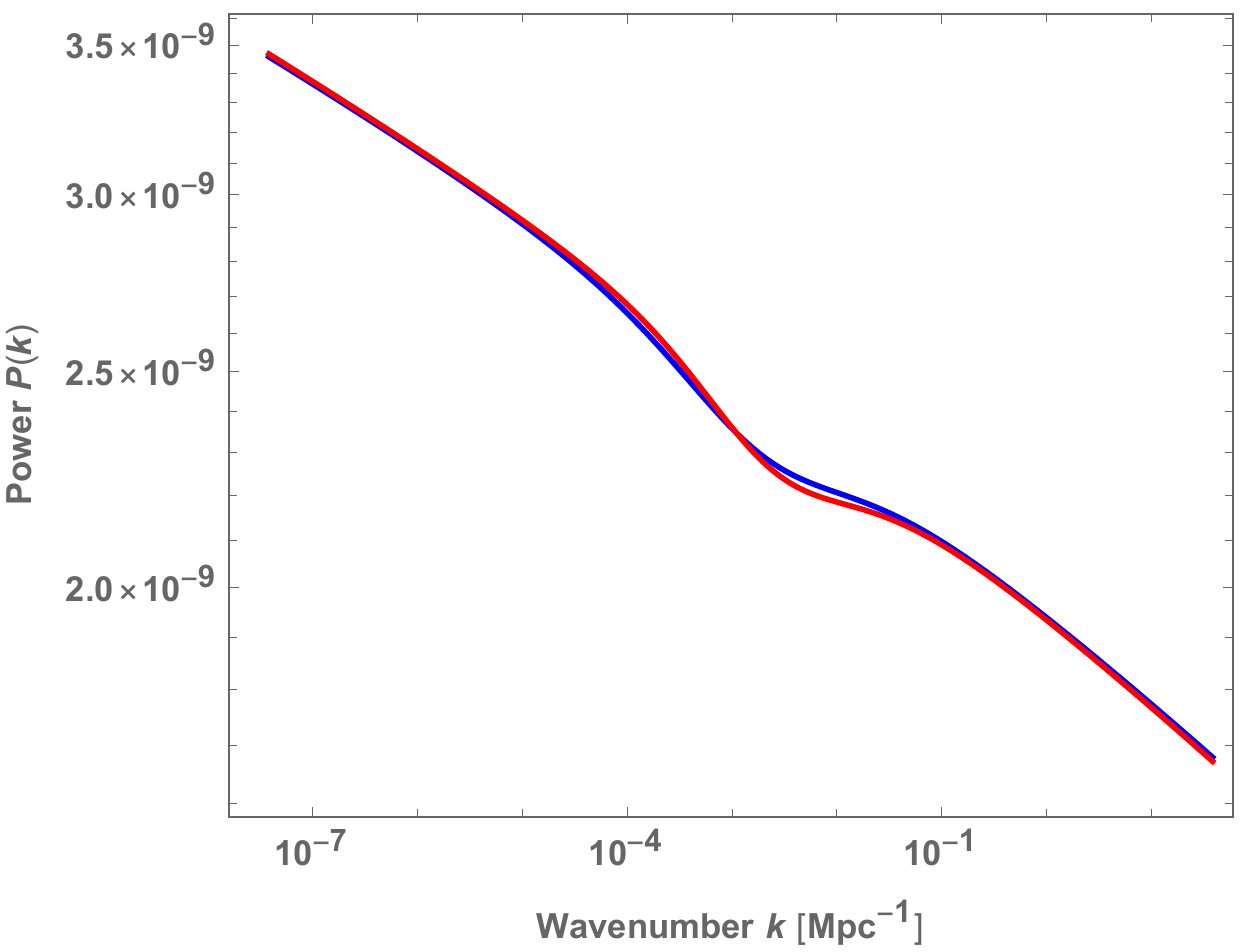}\vspace{0.2cm}\\
\includegraphics[width=8cm]{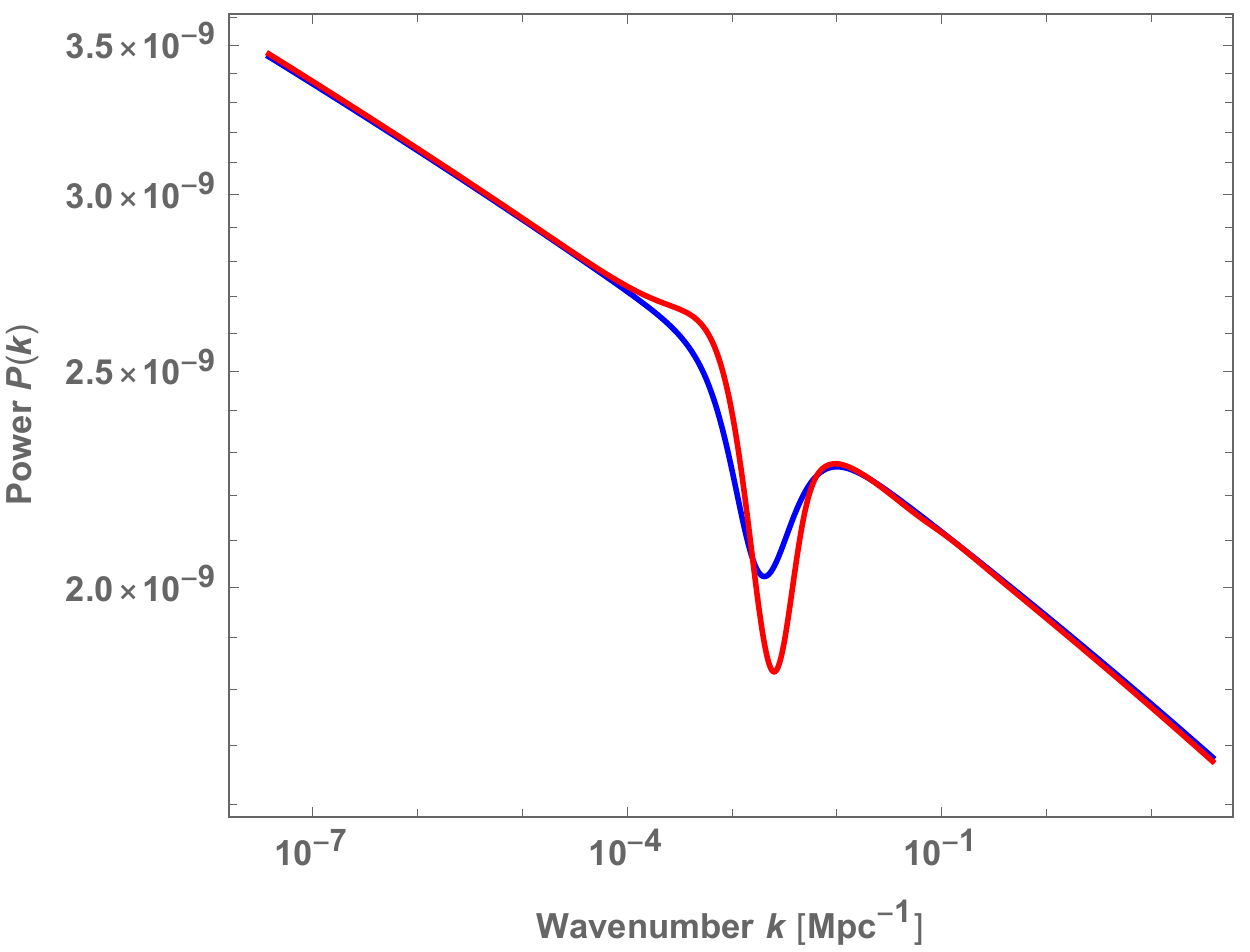}\vspace{0.2cm}\\
\includegraphics[width=8cm]{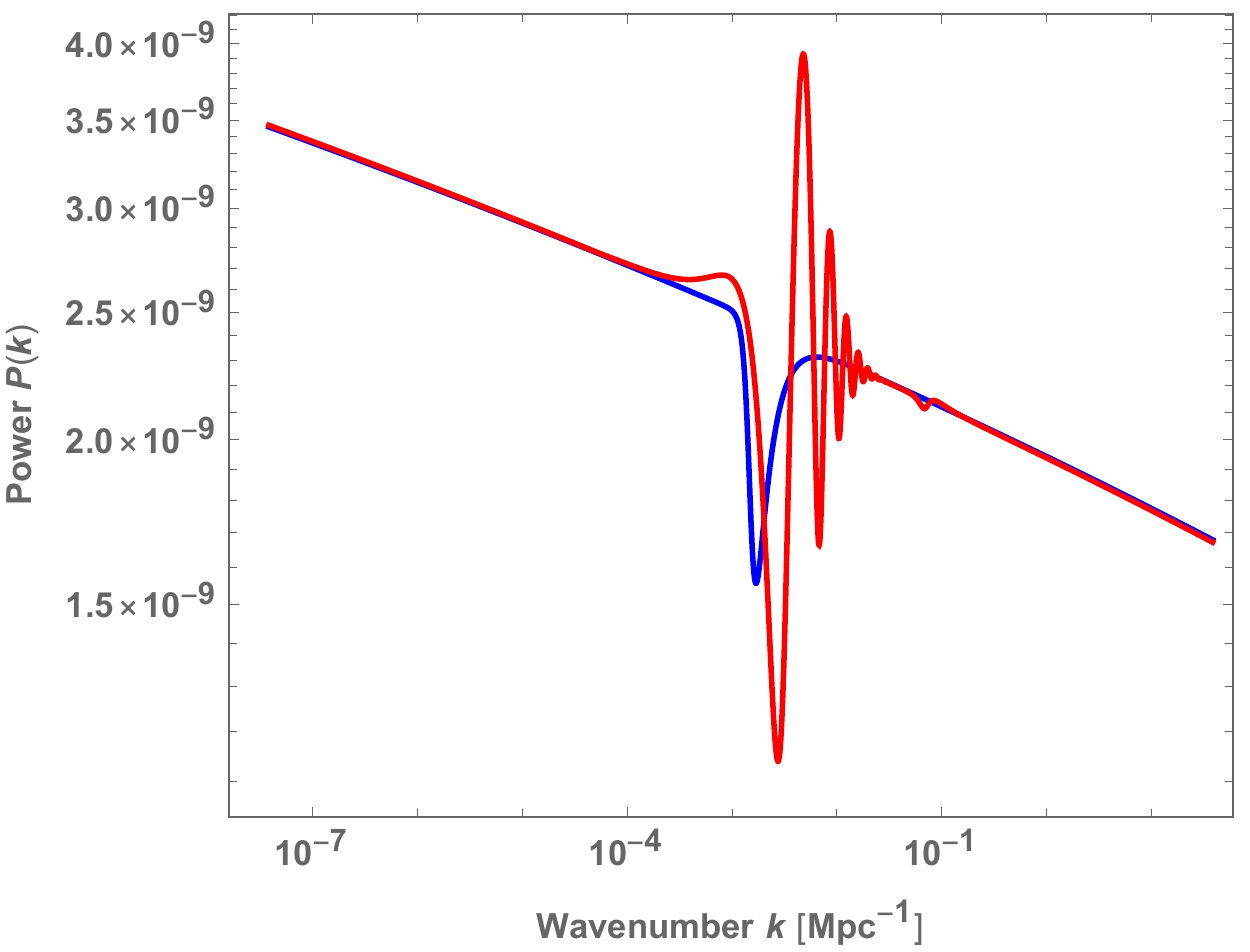}
\caption{The primordial power spectrum ${\cal P}_\zeta(k)$ vs wavenumber $k$ for the potential eqs.~(\ref{V0},\,\ref{deltaV}); the blue curves are from using the approximate method eq.~(\ref{Papprox}) and the red curves are from using the exact method eq.~(\ref{Pexact}). Here we have fixed the parameters $\alpha=0.0015$, $\phi^*=15.243\,\Mpl$, while changing the parameter $\gamma$: From top to bottom $\gamma=3$, $\gamma=10$, and $\gamma=50$, respectively.}
\label{PowerPlot}
\end{figure}

Using the above potential we have obtained the power spectrum in both the approximate and exact methods. In Fig.~\ref{PowerPlot} we give some representative results of this primordial power spectrum. We have introduced the step-like feature at $\phi^*=15.243\,\Mpl$, which is around $N_e\sim 55$ e-foldings before the end of inflation. This will have an imprint on the CMB on large scales, as we analyze in the next section.

The plot shows that as we increase the parameter $\gamma$, that controls the (inverse) width of the step-like feature, the approximate method becomes less and less accurate. Instead the exact numerical method shows that for sufficiently large $\gamma$ the true answer involves significant oscillations. This makes good physical sense for the following reasons: All modes begin in the Bunch-Davies vacuum on sub-horizon scales. In this regime, the modes are oscillating and they carry different phases depending on their wave-number. As the modes red-shift they eventually encounter the feature in the potential. Different modes will encounter the feature in the potential with different phases, and therefore can have different responses; some modes get enhanced and some modes do not, leading to oscillations. Modes then move super-horizon and get frozen (for other discussion, see Ref.~\cite{Kaloper:2003nv}).

We see that the approximate result clearly exhibits the desired dip in the power and is localized in $k$-space, while it can be more complicated when one turns to the exact result. 
There exists some interesting work on computing the spectrum accurately using the ``Generalised Slow Roll" formalism \cite{Stewart:2001cd,Chluba:2015bqa}. However, since the potential may be so sharp, it suffices to do this numerically. We leave it as possible further work to develop an analytical understanding of the resulting behavior.

In the first and second plots in Fig.~\ref{PowerPlot}, one sees that {\em both} the approximate and the exact spectra exhibit this desired suppression. However, since the parameter $\gamma=3$ and $\gamma=10$ in these plots is not very large, the corresponding width of wavenumbers that are affected is appreciable. We shall investigate if any of these types of power spectra are helpful in matching CMB data next.

\section{Predictions for CMB}\label{CMB}

The primordial power spectrum is not directly observable on the largest of scales. Though we can gain information by turning to large scale structure; this is ultimately connected to the primordial spectrum and this mapping is perhaps best quantified in the framework of the effective field theory or large scale structure \cite{Baumann:2010tm,Carrasco:2012cv,Hertzberg:2012qn}. However, the cleanest and most precise information we have on the primordial power spectrum comes directly from the CMB.

In the linear theory it is straightforward in principle to convert the primordial power spectrum ${\cal P}_\zeta(k)$ into a prediction for the fluctuations in the CMB. The primordial spectrum provides the initial conditions for the early radiation and matter dominated eras. This can be evolved using standard plasma and gravitational physics. 

The CMB temperature fluctuation across the 2-sphere $T(\theta,\varphi)$ is decomposed into the spherical harmonic basis $Y_{l,m}(\theta,\varphi)$ as
\beq
{\Delta T(\theta,\varphi)\over \bar{T}} =\sum_{lm}a_{lm}\,Y_{lm}(\theta,\varphi),
\eeq
where in principle an amplitude $a_{lm}$ can be measured for each $l$ and $m$. To approximate the ensemble average of the two-point correlation function $\langle \Delta T(\theta,\varphi)\,\Delta T(\theta',\varphi')\rangle$, a sum over $m=-l,-l+1,\ldots,l-1,l$ is performed for each $l$. The summed power in each squared multipole moment is designated ${\cal D}_l^{TT}$ (up to an overall constant prefactor for convenience) as
\beq
{\cal D}_l^{TT}={l(l+1)\over2\pi(2l+1)}\sum_{m=-l}^{m=l}|a_{lm}|^2.
\eeq

The theoretical prediction for the ensemble average of these multipole moments can be done efficiently numerically, and we make use of the publicly available CLASS program \cite{Blas:2011rf} to carry this out. To do this we have first computed the primordial power spectrum using the above approximate and exact methods. We then inserted this into the CLASS code, running at high precision, and obtained a range of different results.. We used the best fit cosmological parameters taken from Table 3 of Ref.~\cite{Ade:2015xua}, namely
\beq
&&\omega_{b}=0.02222, \,\,\,\omega_{c}=0.1197,\nonumber\\
&&\tau=0.078, \,\,\, H_0=67.31\,\mbox{km}\,\mbox{s}^{-1}\,\mbox{Mpc}^{-1},\,\,\,\,\,\,
\eeq
with current CMB mean temperature of $T_0=2.7255$\,K. 

\begin{figure}[h!]
\centering
\includegraphics[width=7.5cm]{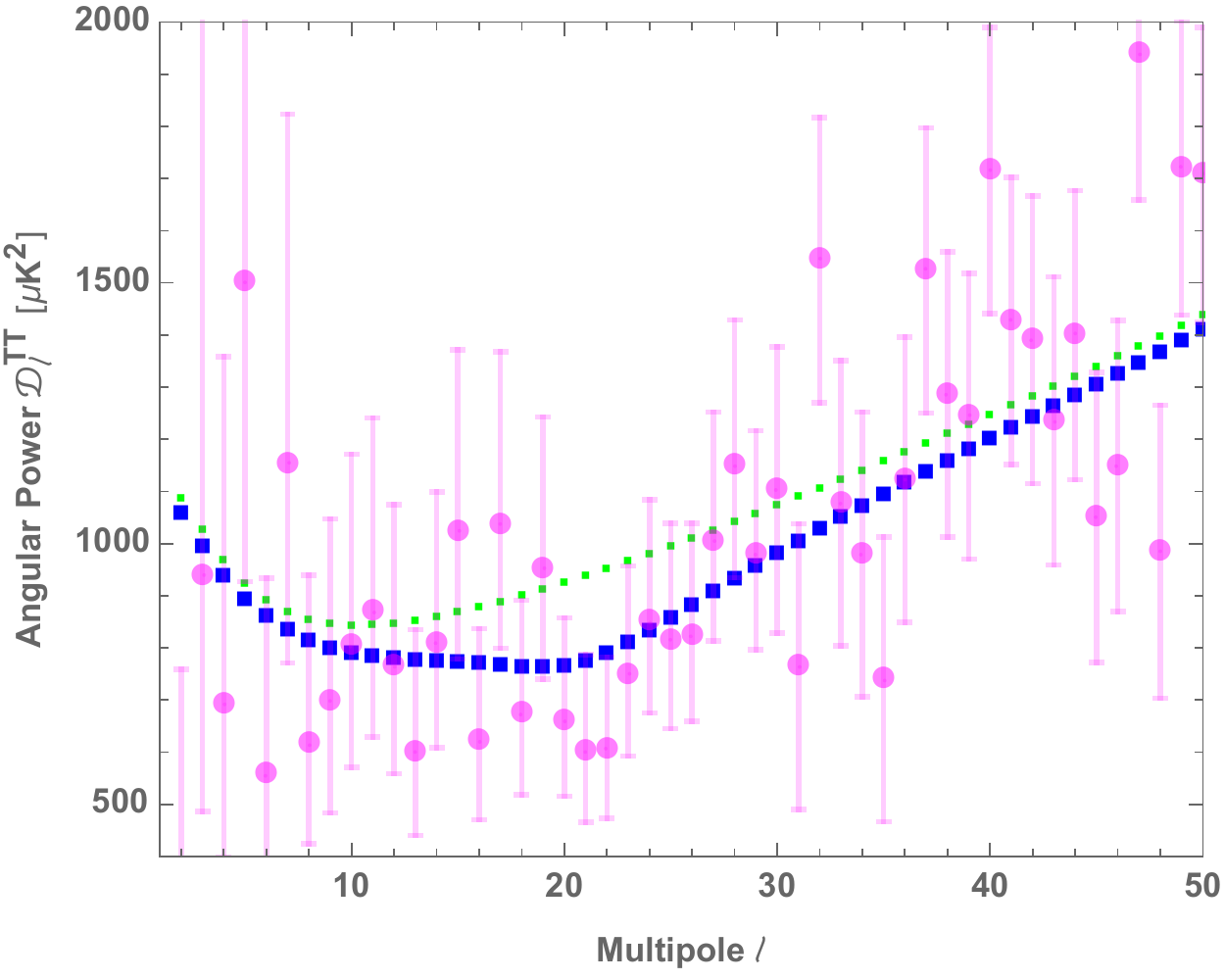}\vspace{0.2cm}\\
\includegraphics[width=7.5cm]{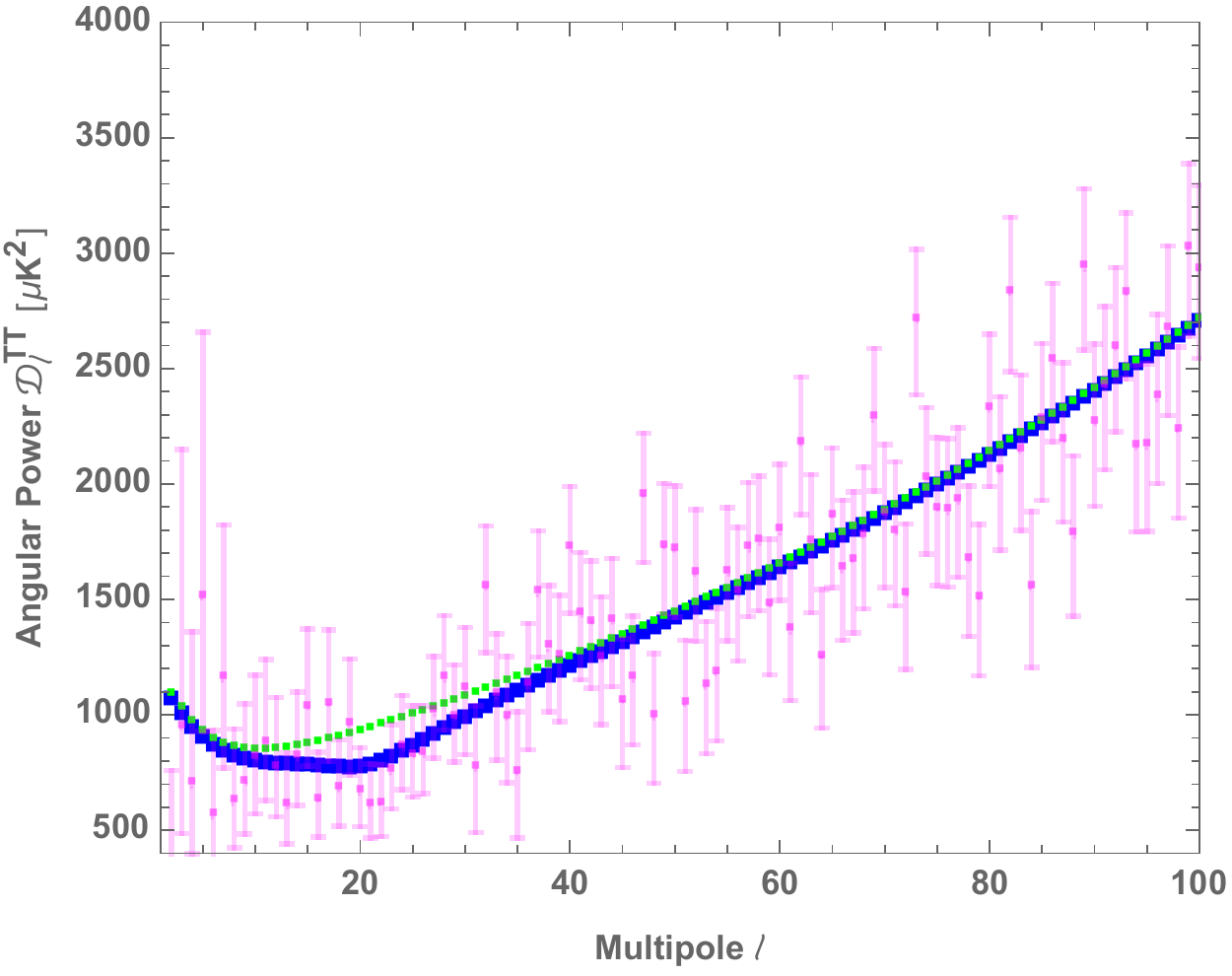}\vspace{0.2cm}\\
\includegraphics[width=7.5cm]{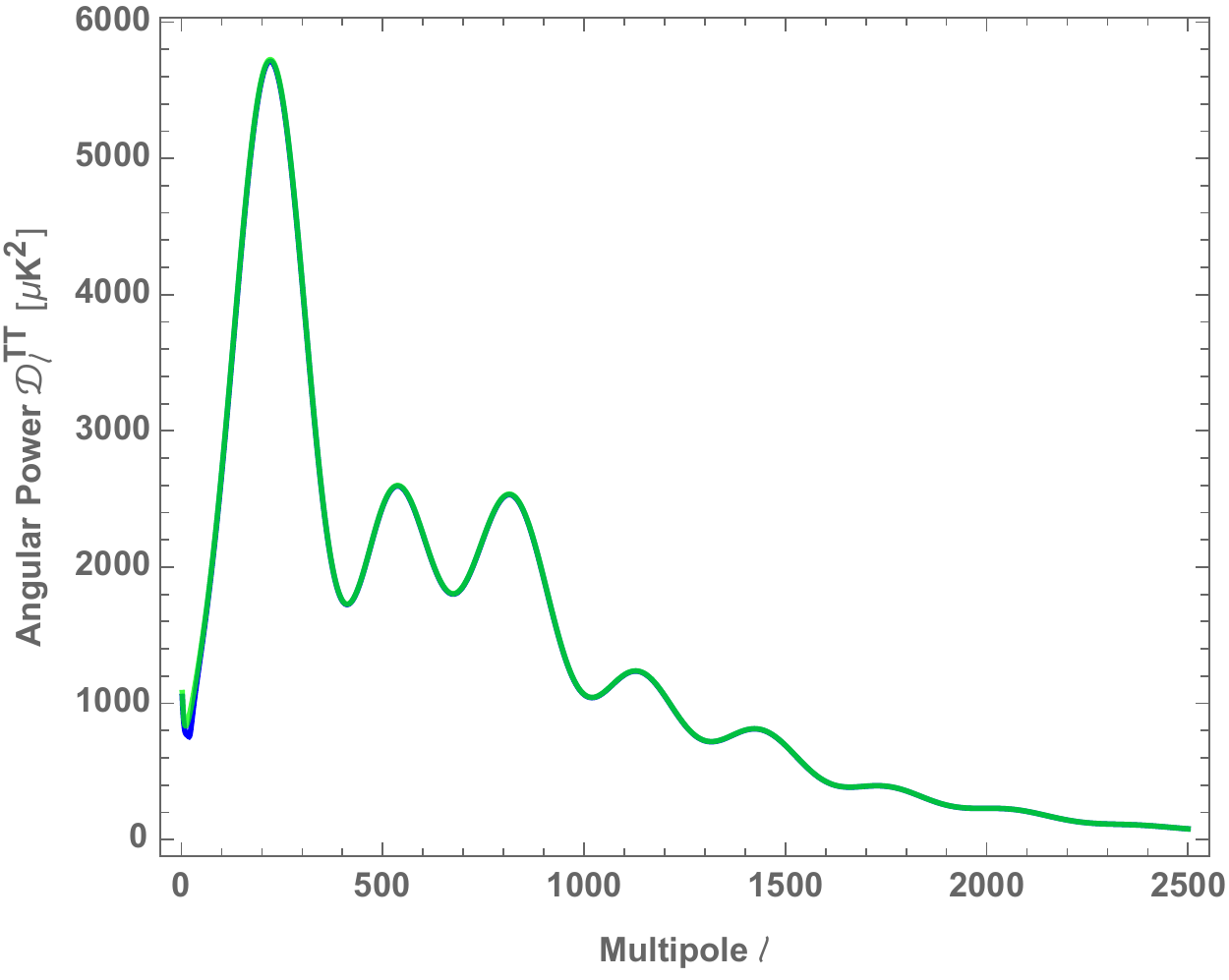}
\caption{The CMB multipole moments ${\cal D}_l^{TT}$ versus multipole $l$ using the {\em approximate} power spectrum from the slow-roll approximations with potential $V$ (blue points). Here $\alpha=0.0015$, $\gamma=50$, and $\phi^*=15.243\,\Mpl$. We also show the nearly scale invariant model with potential $V_0=m^2\phi^2/2$ (green points) and Planck data \cite{PlanckLegacy} (magneta points). Top panel is $2\le l \le 50$, middle panel is $2\le l\le 100$, and bottom panel (where we suppress the Planck data) is $2\le l \le 2500$.}  
\label{MultipolePlotApprox}
\end{figure}

In Fig.~\ref{MultipolePlotApprox} we give a representative plot of the CMB multipole moments using the approximate method. We have chosen parameters that coincide with those of the bottom plot of Fig.~\ref{PowerPlot}. The parameters are chosen such that we obtain a clear dip in the spectrum (blue points) compared to the nearly scale invariant theory with potential $V_0=m^2\phi^2/2$ (green points). This shows reasonable agreement with the Planck data \cite{PlanckLegacy} (magenta points), and in particular improves agreement with the suppression in power around $l\sim 20-30$ as desired.

\begin{figure}[t!]
\centering
\includegraphics[width=7.6cm]{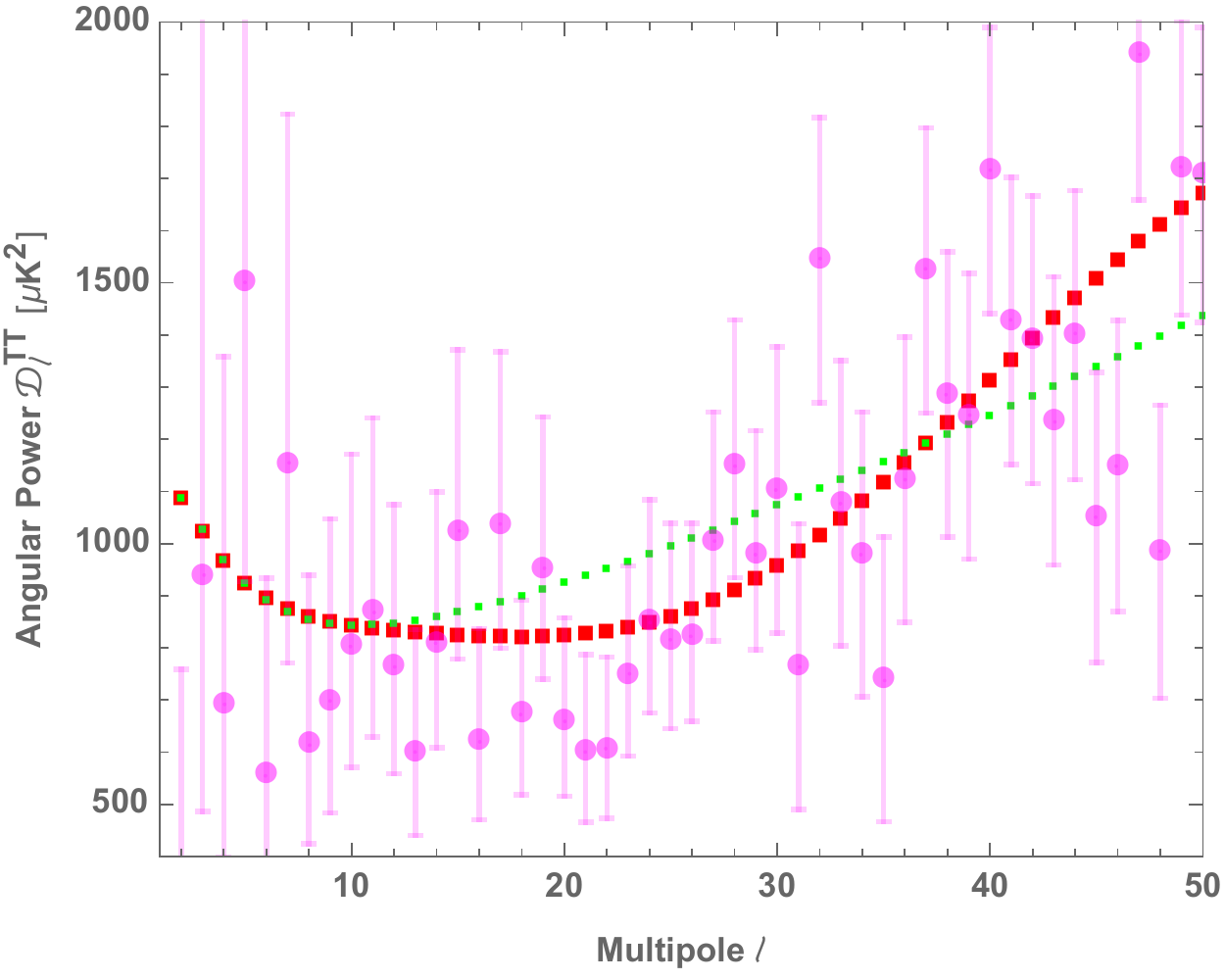}\vspace{0.2cm}\\
\includegraphics[width=7.6cm]{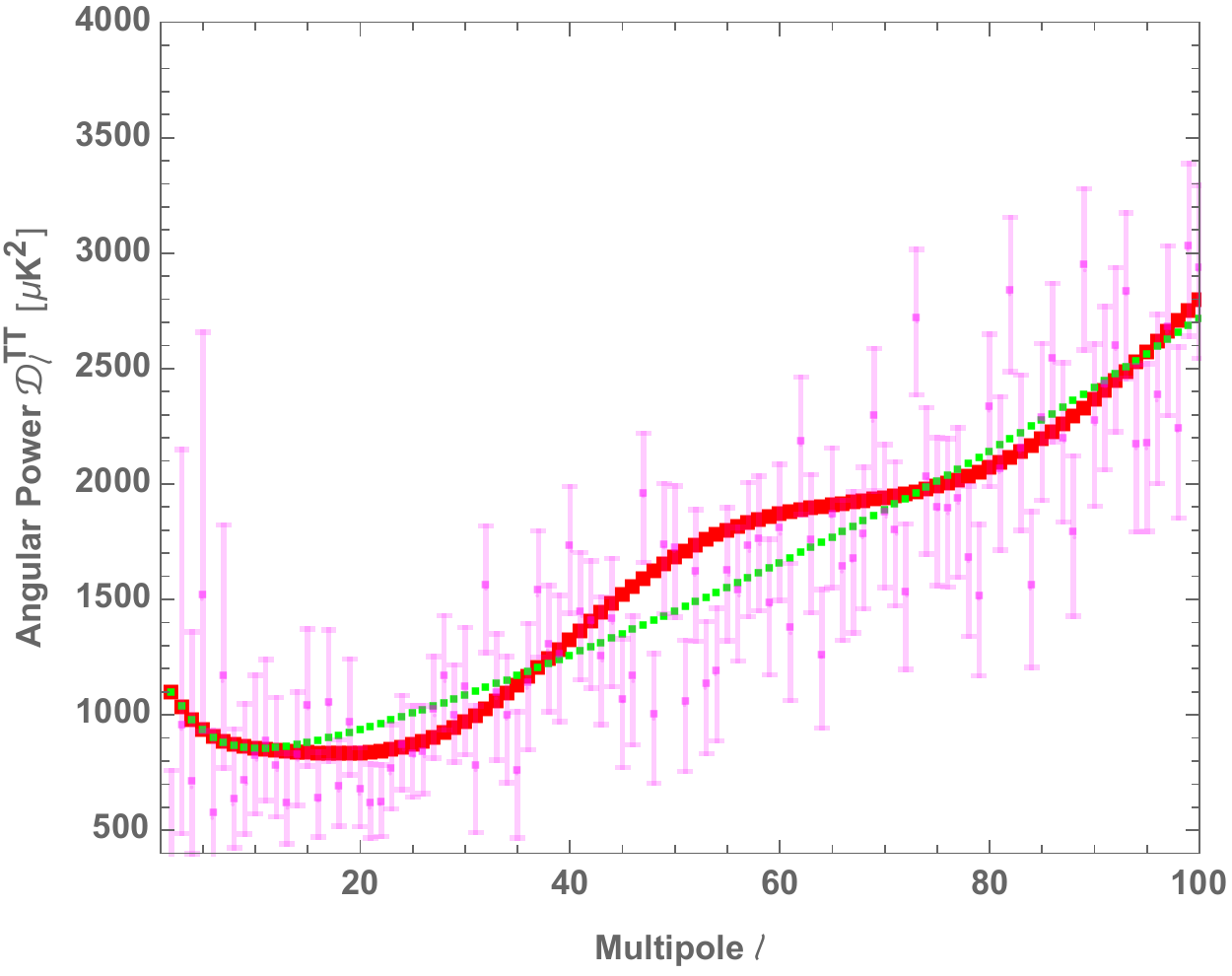}\vspace{0.2cm}\\
\includegraphics[width=7.6cm]{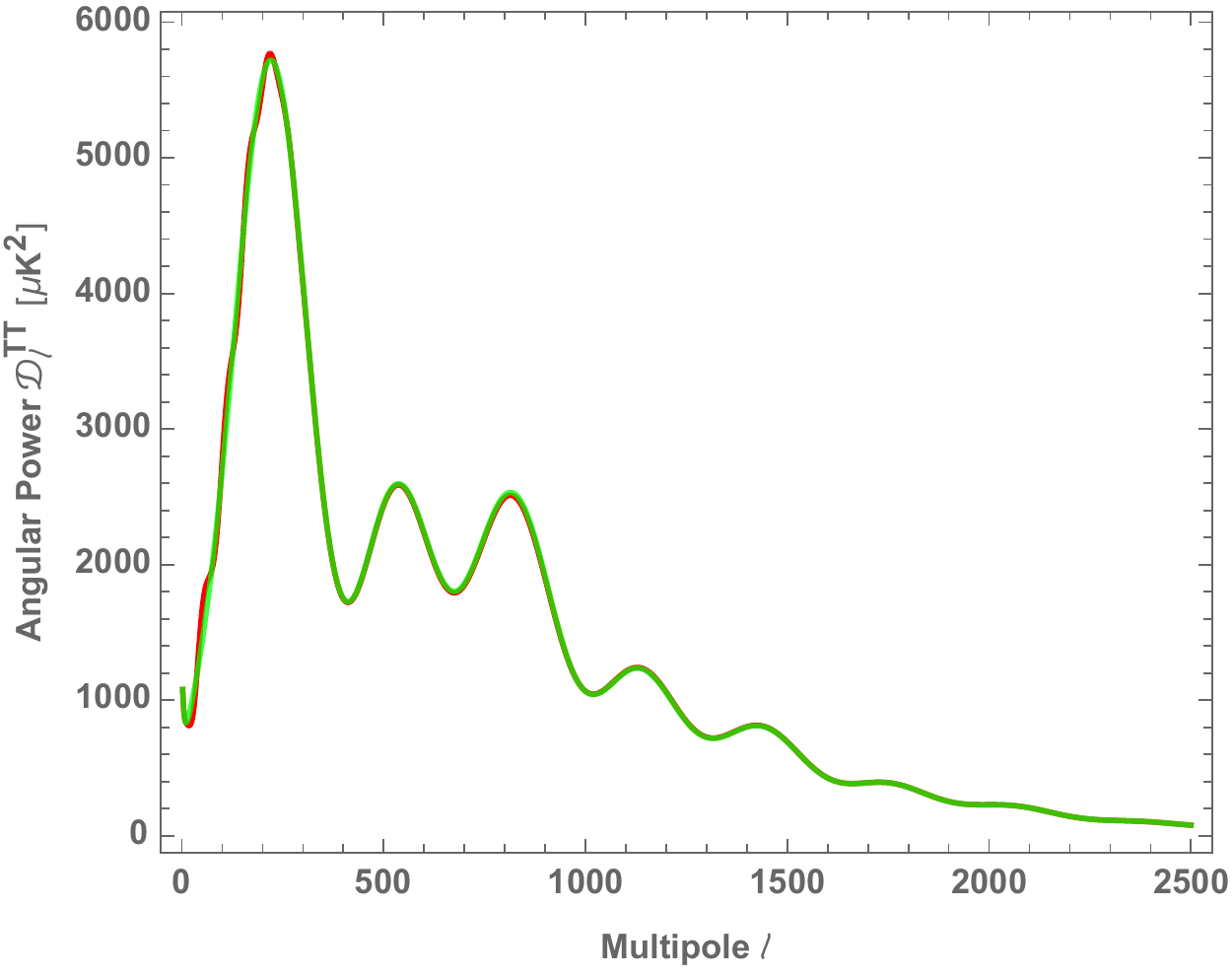}
\caption{The CMB multipole moments ${\cal D}_l^{TT}$ versus multipole $l$ using the {\em exact} power spectrum from solving for the MS variable numerically with potential $V$ (red points). The parameters are the same as in Fig.~\ref{MultipolePlotApprox}.}
\label{MultipolePlotExact}
\end{figure}

However, as was to be anticipated from Fig.~\ref{PowerPlot}, the results using the exact method are rather different. The presence of multiple oscillations in the primordial power spectrum translates into oscillations in the CMB multipoles. In fact whenever we make $\gamma$, the (inverse) width, sufficiently large to attempt to localize the dip using the approximate method, we find that this gives rise to a break down in the slow-roll condition $\epsilon_2\ll 1$ and the exact method exhibits poor agreement with data, as seen in Fig.~\ref{MultipolePlotExact}. This was to be anticipated from Fig.~\ref{SlowRoll}. However, in order to quantify the comparison between theory and data we now turn to a statistical analysis.
\newpage

\section{Statistical Analysis}\label{Statistics}

A measure of the difference between theory and data is the sum of squares of the differences, normalized to the variance. This is the so-called $\chi^2$ statistic. It counts the number of degrees of freedom in the case in which we have the correct theory. Here we can define it as a sum over multipoles $l$ as
\be
\chi^2 \equiv \sum_{l=l_{\rm min}}^{l_{\rm max}}{({\cal D}_{l,{\rm theory}}^{TT}-{\cal D}_{l,{\rm data}}^{TT})^2\over\sigma_l^2},
\ee
where the first factor of ${\cal D}_{l,{\rm theory}}^{TT}$ refers to the theoretical prediction, using either the approximate or exact method, and ${\cal D}_{l,{\rm data}}^{TT}$ refers to the measured central value of the Planck data. The variance is in general a combination of the theoretical uncertainty $\sigma_{l,{\rm theory}}^2$, since inflation is a statistical theory based on quantum mechanics, and statistical uncertainty $\sigma_{l,{\rm data}}^2$, due to the fact that detectors are imperfect, the presence of foregrounds, and cosmic variance. For concreteness we take this factor in the denominator to be
\be
\sigma_l^2 = \mbox{Max}\left\{\sigma_{l,{\rm theory}}^2,\sigma_{l,{\rm data}}^2\right\},
\ee
with theoretical variance
\be
\sigma_{l,{\rm theory}}^2={2\over 2l+1}{\cal D}_{l,{\rm theory}}^2,
\ee
and statistical variance $\sigma_{l,{\rm data}}^2$ is read off from the reported Planck error bars. The multipole moments begin from $l_{\rm min}=2$ and we go up to $l_{\rm max}=2500$.

\begin{figure}[t!]
\centering
\includegraphics[width=10cm]{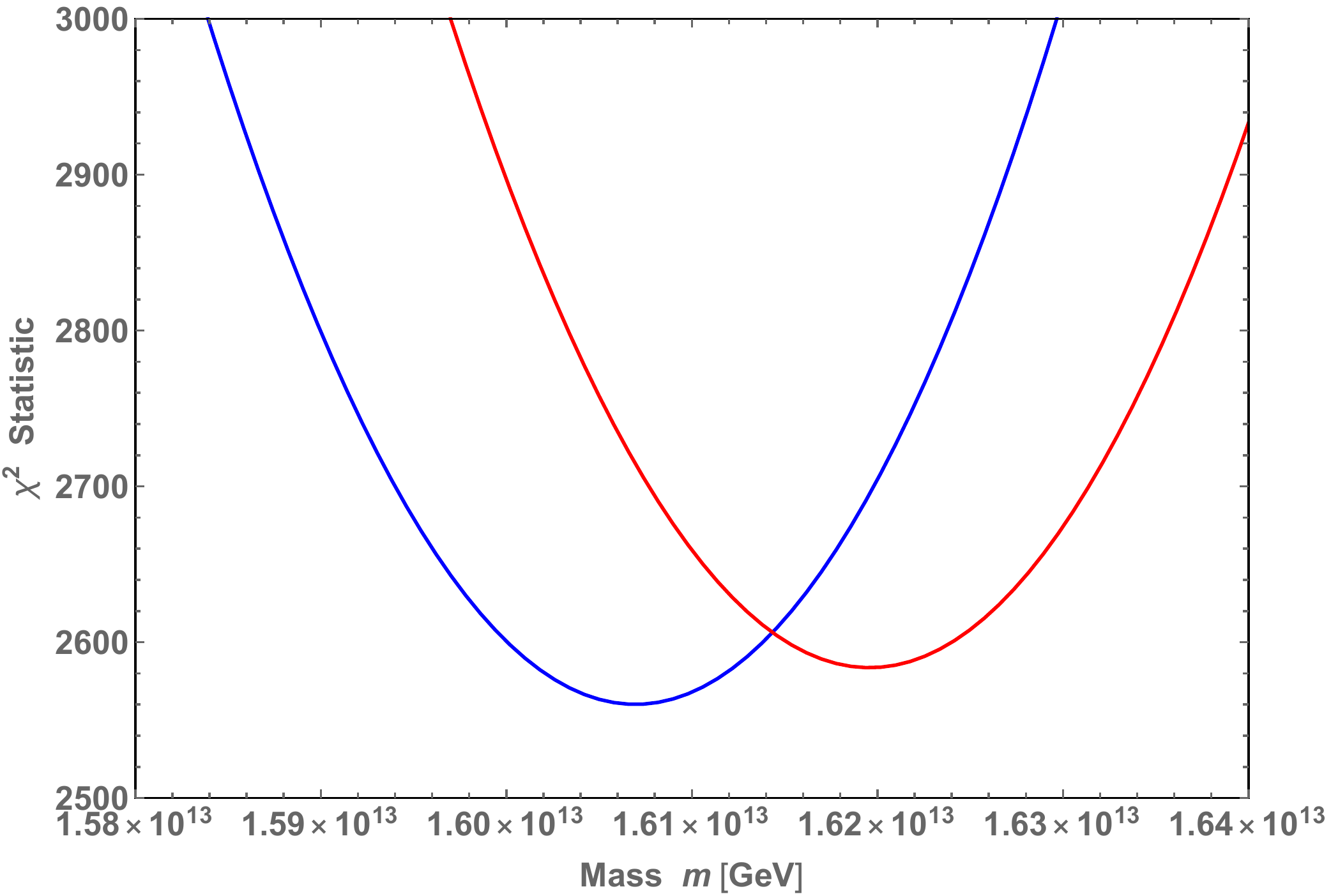}
\caption{A statistical measure of the comparison between theory and data $\chi^2$ as a function of the mass parameter of the potential $m$, with $\alpha=0.0015$, $\gamma=50$, and $\phi^*=15.243\,\Mpl$. The blue curve is from using the approximate method and the red curve is from using the exact method. The optimal $m$ is the one that minimizes this function.}
\label{StatisticsMassPlot}
\end{figure}

We need to choose a value of the inflaton mass $m$ in order to specify our model. As is true of essentially {\em any} inflationary model this overall scale of the potential needs to be optimized to fit the data; as there is no known microphysics that determines $m$. We have selected $m$ by optimizing our $\chi^2$ statistic. A plot of $\chi^2$, for fixed $\alpha$, $\gamma$, and $\phi^*$, as a function of $m$ is given in Fig.~\ref{StatisticsMassPlot}. The optimal choice of $m$ is the one that minimizes $\chi^2$.

\begin{figure}[t!]
\centering
\includegraphics[width=10cm]{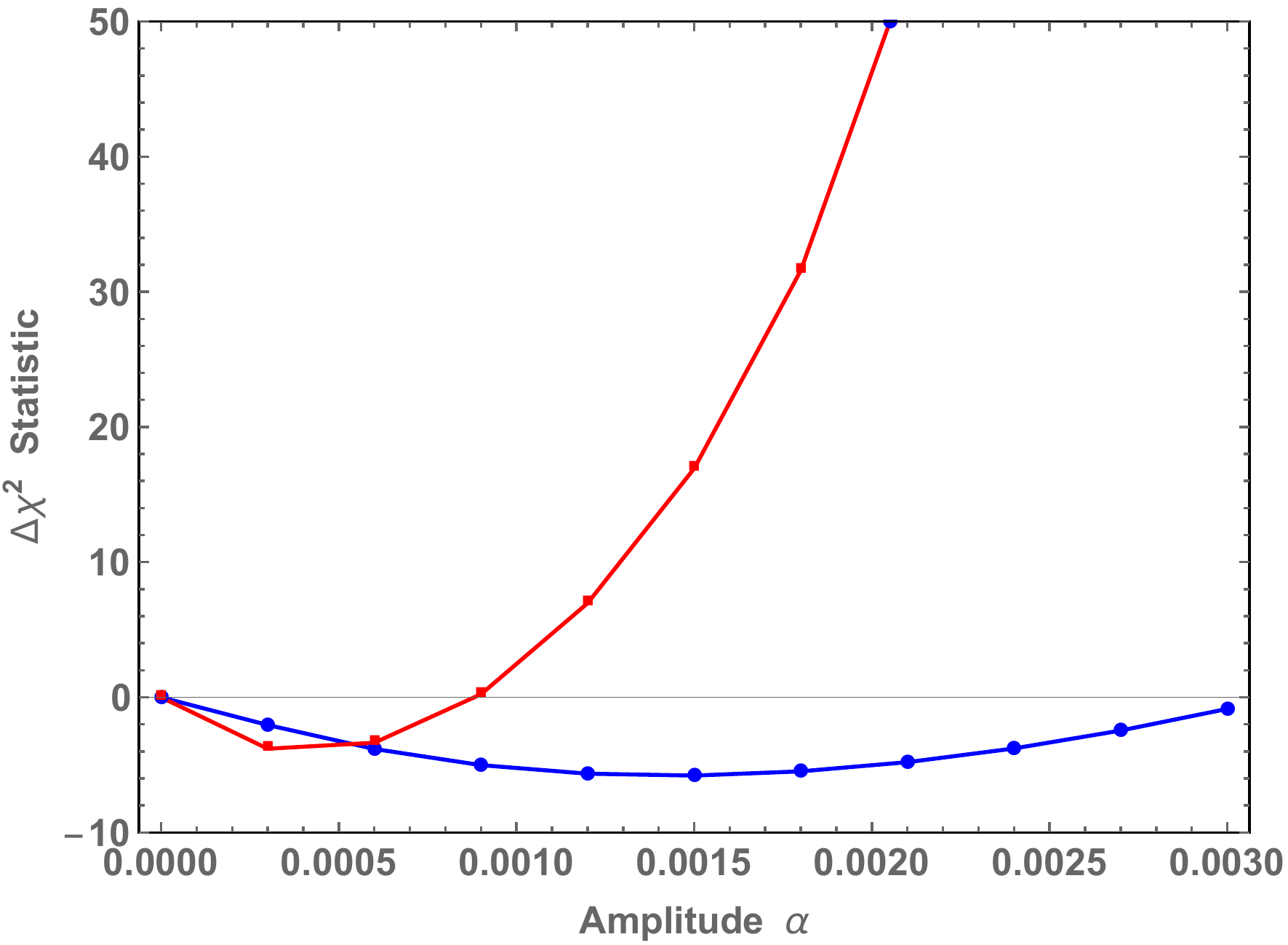}\vspace{0.4cm}\\
\includegraphics[width=10cm]{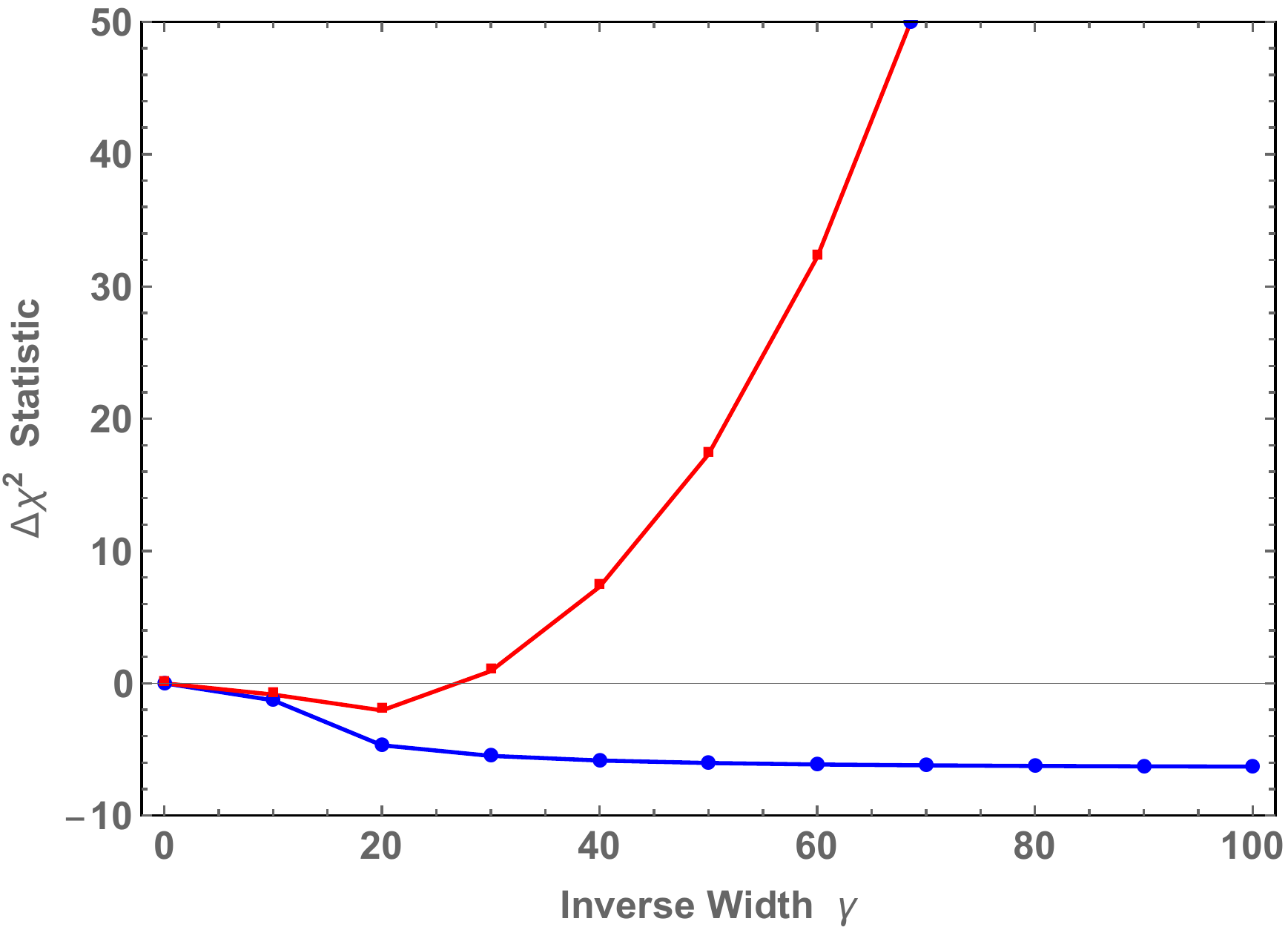}
\caption{A statistical measure of the comparison between theory and data $\Delta \chi^2=\chi^2-\chi_0^2$ in the potential with a feature versus the featureless potential as a function of the parameters in the model. Top panel: we vary $\alpha$ at fixed $\gamma=50$ and $\phi^*=15.243\,\Mpl$. Bottom panel: we vary $\gamma$ at fixed $\alpha=0.0015$ and $\phi^*=15.243\,\Mpl$. The blue points are from using the approximate method and the red points are from using the exact method.}
\label{StatisticsPlot}
\end{figure}

Having optimized for the overall scale $m$, we turn to how our statistic varies as we change the shape parameters of the potential. Since our interest is the {\em change} in our statistical measure $\chi^2$ in the new theory with the step-like feature $\delta V$ from the standard potential without it $\chi_0^2$, we will in fact report on the difference 
\be
\Delta \chi^2 \equiv \chi^2 -\chi_0^2.
\ee
In Fig.~\ref{StatisticsPlot} we show the value of $\Delta \chi^2$ as a function of $\alpha$ (top panel) and as a function of $\gamma$ (bottom panel). We observe that using the approximate method (based on the assumption of slow-roll) there is some moderate reduction in the value of $\chi^2$ (as seen in the negative values of $\Delta \chi^2$). This is in accord with what one can see by eye in Fig.~\ref{MultipolePlotApprox}; the presence of the dip in the power spectrum leads to a suppression in power in the multipole moments in just the right place to improve agreement with data around $l\sim 20-30$. The reduction in $\chi^2$ compared to the standard theory $\chi_0^2$ is at most around $\sim 7$ (i.e., $\Delta \chi^2\sim-7$). We note that this is larger than the number of new parameters in the model of 3 for $\{\alpha,\,\gamma,\,\phi^*\}$ (or effectively only 2 as we make $\gamma$ very large). We do not claim that this is highly significant, nevertheless according to the slow-roll approximation, a moderate improvement in $\chi^2$ is achievable.

However, when we turn to the exact method (which does not assume slow-roll) we see the situation is much worse. Generally as we increase both $\gamma$ and $\alpha$ to large values, the theory does a much worse job in fitting the data than the standard nearly scale invariant theory, as seen in the significant growth of $\Delta \chi^2$ in Fig.~\ref{StatisticsPlot}. This was to be expected from Fig.~\ref{MultipolePlotExact} in which it was seen that large oscillations occur in the multipole moments, which are not seen in the data in this fashion. We note that for small values of $\alpha$ and $\gamma$, we do obtain $\Delta \chi^2<0$. However the best we achieve is $\Delta \chi^2\sim -3$, which is comparable to the number of new parameters in the model of 3. So overall there is really no improvement when the exact computation is performed. In general we believe that this type of problem will likely persist for any relatively simple model of inflation that attempts to explain the suppression in the data.

\section{Discussion}\label{Discussion}

We have shown that models which provide a suppression in power on the appropriate scales arise from a potential that has a step-like feature. However, to provide a significant improvement, this feature needs to be so sharp to that the standard slow-roll approximation for the power spectrum breaks down and an exact numerical approach is required. In this case the same potential functions in fact lead to rapid oscillations in the spectrum and affects other scales too, which does not fit the data well. 

A possible future approach is to ``reverse engineer" the potential $V$, by instead starting with the data, and constructing a potential function that can reproduce it. This would be similar to the idea of Ref.~\cite{Hertzberg:2017dkh} that did this in order to construct an appropriate spike in the matter power spectrum leading to primordial black holes. We anticipate, however, that in order to obtain a 
localized suppression in power without the large oscillations, the corresponding potential $V$, if it exists, will have an extremely peculiar shape. It must be of a very special form for all these oscillations to conspire to cancel out among the various features of the potential. One may attempt to use a potential which itself has an oscillatory feature that may give rise to a sharp feature upon Fourier transforming, but for this to extend to the CMB multipoles in just the right way, appears rather difficult (interesting work includes Ref.~\cite{Zeng:2018ufm}). We also note that models with non-trivial features in the potential may also give rise to significant non-Gaussianity, which could also rule them out for independent reasons, so this is another important constraint to satisfy.

This suggests that a potential function that leads to just the desired feature of a dip in the spectrum would likely be highly fine-tuned from the microscopic point of view. On the other hand, an overall suppression in power in all low $l$-modes may be possible. However, we leave a detailed exploration for future work.

\section*{Acknowledgments}

MPH is supported by National Science Foundation grant PHY-1720332, a Gordon Godrey fellowship, a JSPS fellowship, and this work was supported in part by a Tufts FRAC award.

\end{document}